# Initial Sampling in Symmetrical Quasiclassical Dynamics Based on Li-Miller Mapping Hamiltonian

Jie Zheng[1,2], Yu Xie[3,4]*, Shengshi Jiang[5,6], Yunze Long[1,2], Xin Ning[1], Zhenggang Lan[3,4]*

[1]Industrial Research Institute of Nonwovens & Technical Textiles, College of Textiles Clothing, Qingdao University, Qingdao 266071, China

[2]College of Physics, Qingdao University, Qingdao 266071, China

[3]SCNU Environmental Research Institute, Guangdong Provincial Key Laboratory of Chemical Pollution and Environmental Safety & MOE Key Laboratory of Theoretical Chemistry of Environment, South China Normal University, Guangzhou 510006, China

[4]School of Environment, South China Normal University, University Town, Guangzhou 510006, China

[5]Qingdao Institute of Bioenergy and Bioprocess Technology, Chinese Academy of Sciences, Qingdao 266101, China

[6]University of Chinese Academy of Sciences, Beijing 100049, China

* E-mail：zhenggang.lan@m.scnu.edu.cn, zhenggang.lan@gmail.com(ZL); yu.xie@m.scnu.edu.cn (YX).

# Abstract

A symmetrical quasiclassical (SQC) dynamics approach based on the Li-Miller (LM) mapping Hamiltonian (SQC-LM) was employed to describe nonadiabatic dynamics. In principle, the different initial sampling procedures may be applied in the SQC-LM dynamics, and the results may be dependent on the initial sampling. We provided various initial sampling approaches and checked their influence. We selected two groups of models including site-exciton models for exciton dynamics and linear vibronic coupling models for conical intersections to test the performance of SQC-LM dynamics with the different initial sampling methods. The results were examined with respect to those of the accurate multilayer multiconfigurational time-dependent Hartree (ML-MCTDH) quantum dynamics. For both two models, the SQC-LM method more-or-less gives a reasonable description of the population dynamics, while the influence of the initial sampling approaches on the final results is noticeable. It seems that the proper initial sampling methods should be determined by the system under study. This indicates that the combination of the SQC-LM method with a suitable sampling approach may be a potential method in the description of nonadiabatic dynamics.

# I. INTRODUCTION

The simulation of nonadiabatic processes in large complex systems is challenging due to the breakdown of the Born-Oppenheimer approximation and the involvement of a large number of degrees of freedom. [1-3] Several advanced dynamical approaches, such as full quantum dynamics methods, [4, 5] Gaussian-wavepacket-based dynamics [6-15] and rigorous semiclassical dynamics methods [1, 2, 16-19], are computationally expensive. Although a few of the practical methods such as Ehrenfest dynamics [20] (or its extension [21]) and trajectory surface hopping methods [16, 22-28] are popular due to their computational efficiency, their deficiencies are also well known [16, 24, 25, 29-32].

In 1979, Meyer and Miller [33] constructed a mapping Hamiltonian in the description of nonadiabatic dynamics. Later, this idea was further rederived by Stock and Thoss [34] by using Schwinger's transformation. In short, the MM (Meyer-Miller) mapping model constructs a mapping protocol from $N$ discrete quantum states to $N$ coupled harmonic oscillators. Previous works showed that it is possible to employ a mapping Hamiltonian in the treatment of nonadiabatic dynamics. [16, 17, 35-57]

In the original paper describing the MM mapping Hamiltonian [33], Meyer and Miller discussed that it was necessary to introduce the zero-point (ZP) energy of the harmonic oscillator into the mapping Hamiltonian. Without such a term, the classical dynamics based on the classical mapping Hamiltonian gives the Ehrenfest dynamics. In the later work by Stock and Thoss [34], they also obtain the ZP energy term in the construction of the mapping Hamiltonian. According to results from numerical testing, Previous works recommended to include only a factional part of the zero-point energy (ZPE) [55-57] for the electronic mapping degrees of freedom in the classical treatment based on the MM mapping model. Recently, Miller and coworkers proposed a symmetrical quasi-classical dynamics method based on the MM mapping Hamiltonian (SQC-MM)[58-60]. This method received considerable attention [58, 60-72] due to its good performance in terms of both efficiency and accuracy in the treatment of several typical nonadiabatic systems. Recently, several SQC-MM related techniques[66, 73-78], such as the triangle windowing technique for weak-nonadiabatic-coupling cases[66, 79], the modified SQC-MM method for systems with anharmonicity [74], the multistate trajectory approach[75, 76] and the trajectory-adjusted electronic ZPE technique were developed[80].

Miller and coworkers have pointed out that it is also possible to construct the mapping models in other ways[47, 81-83]. For instance, Cotton and Miller [82] showed that the mapping Hamiltonian could be constructed by mapping each electronic state to a spin ½ degree of freedom rather than a harmonic oscillator. In their benchmark calculations, the results showed that these two classical mapping Hamiltonians gave very similar results in most cases, while in other cases, an obvious difference existed. Overall, the MM mapping version always gave somewhat better results.

Alternatively, Li and Miller proposed to construct the mapping model by mapping several electronic states to several coupled angular momenta[81]. The $z$-axis component of quantum angular momentum is mapped to its classical counterpart expressed by spatial coordinates and momenta on the *x*- and *y*-axes. In their work, they also show that the dynamics based on such mapping Hamiltonian with the semiclassical initial value representation (SC-IVR) gives the promising results in the many-electron fermion systems. [81] Liu [84] proposed a unified theoretical framework to derive the mapping model, in which creation/annihilation operators were introduced, and their commutation/anti-commutation relations were derived in a rather rigorous way. Based on these basic relations, several mapping models can be derived by the consideration of various physical variables, which includes the MM and LM models. In the recent work by Liu[85], the LM model was combined with the linearized SC-IVR (LSC-IVR) approach to treat nonadiabatic dynamics in the electron-nucleus coupled models, giving the promising results.

In principle, different mapping Hamiltonians should be equivalent to the original Hamiltonian in the quantum mechanical framework. Thus, in the ideal case, we always wish that the different mapping Hamiltonians can capture and even predict the same dynamical evolution governed by the original Hamiltonian. However, when the classical approximation is introduced in the dynamical treatment, the situation becomes more complicated. Liu[84] already pointed out that the classical evolutions based on the MM and LM mapping Hamiltonians give the same results if the particular initial conditions are considered in the classical dynamics based on LM mapping model. In the MM mapping Hamiltonian, the *N* states are mapped to the *N* coupled harmonic oscillator and each harmonic oscillator is characterized by the coordinate *x* and momentum *p*. In the LM mapping Hamiltonian, the *N* states are mapped to the *N* coupled angular momenta in the *z*-axis and each angular momentum is characterized by four variables, namely two spatial coordinates and two momentum components ($x^{(k)}$, $p_y^{(k)}$, $y^{(k)}$ and $p_x^{(k)}$) in the Cartesian coordinate. Liu[84] noticed that if the initial conditions satisfy the constraint of $x^{(k)}(0) = p_y^{(k)}(0)$ and $y^{(k)}(0) = -p_x^{(k)}(0)$, the classical dynamics based on the MM and LM Hamiltonian is identical in principle. This nicely provides the interesting connection between the classical treatments based on these two mapping Hamiltonian. However, the above constraint in the initial sampling is not a necessary condition. Other initial conditions may also be chosen in the classical treatment based on the LM Hamiltonian. Because there are four variables existing for each angular momentum, some flexible choice of the initial conditions should be possible. In principle, different initial conditions should give different dynamics results.

In this work, we first introduced the symmetrical window approach in the initial sampling and final assignment of the quantum states for the mapping degrees of freedom, giving the symmetrical quasiclassical dynamics for the LM model (SQC-LM). Second, we try to suggest different initial sampling ways in the SQC-LM dynamics. This show the influence of the initial sampling on the final dynamics results. In such way, we systemically evaluate the performance of the SQC-LM method. Here, we mainly focus on two groups of typical model systems. The first group of examples includes symmetric and asymmetric site-exciton models that are widely used in the description of photoinduced exciton dynamics in biological and material systems. We took several site-exciton models and relevant ML-MCTDH results from our recent work [86], which focused on the performance of the SQC-MM method. The second group of models focus on the linear-vibronic coupling models in the description of the nonadiabatic dynamics at conical intersections of polyatomic systems. Here, we took the well-known 3-mode and 24-mode linear vibronic models for the description of the excited-state dynamics of pyrazine.[87, 88] This work should improve our understanding of the theoretical insight for the LM mapping Hamiltonian in the treatment of nonadiabatic dynamics.

## II. THEORETICAL METHODOLOGY

**A. Hamiltonian**

A.1. Site-exciton models

Site-exciton models have been widely used to describe the photoinduced excitonic processes[68, 86, 89-92]. In this paper, we took several site-exciton models from our recent work [86], which represent two localized electronic states (system) coupled to many vibrational modes (bath). The system−bath Hamiltonian $H$ reads

$$H = H_s + H_b + H_{sb}, \tag{1}$$

where $H_s$ denotes the system Hamiltonian (electronic part), $H_b$ is the bath part (vibrational part) and $H_{sb}$ defines the system−bath couplings (electronic-phonon interaction). The system Hamiltonian is

$$H_s = \sum_k |\phi_k\rangle E_k \langle\phi_k| + \sum_{l\neq k} |\phi_k\rangle V_{kl} \langle\phi_l|, \tag{2}$$

where $\phi_k$ ($\phi_l$) represents the $k$-th ($l$-th) electronic state, $E_k$ ($E_l$) represents the site energy of the localized electronic state and $V_{kl}$ represents the interstate coupling between the $k$-th and $l$-th electronic state. Each electronic state couples with an individual bath. The bath Hamiltonian reads

$$H_b = \sum_k \sum_j^{N_b} \frac{1}{2}\omega_{kj} \left[P_{kj}^2 + Q_{kj}^2\right], \tag{3}$$

where $N_b$ is the bath-mode number. $P_{kj}$ and $Q_{kj}$ stand for the momentum and position of the $j$-th vibrational mode in the $k$th bath coupled with the $k$th electronic state. $\omega_{kj}$ is the frequency of the corresponding mode. The electronic-phonon interaction Hamiltonian is

$$H_{sb} = \sum_k |\phi_k\rangle\langle\phi_k| \sum_j^{N_b} \left(-\kappa_{kj} Q_{kj}\right), \tag{4}$$

where $\kappa_{kj}$ is the system-bath coupling constant of the corresponding mode in the diagonal elements of the system Hamiltonian. The off-diagonal elements in the system Hamiltonian do not depend on the system−bath couplings.

The continuous Debye-type spectral density [3] was employed to represent the bath,

$$J(\omega) = \frac{2\lambda\omega\omega_c}{\omega^2 + \omega_c^2}, \tag{5}$$

where $\omega_c$ denotes the characteristic frequency of the bath and $\lambda$ is the reorganization energy. The spectral density can also be defined by using many discrete bath modes [3]

$$J_k(\omega) = \frac{\pi}{2} \sum_{i=1}^{N} \kappa_{ki}^2 \delta(\omega - \omega_{ki}),  \qquad (6)$$

where the electron-phonon coupling $\kappa_{ki}$ is

$$\kappa_{ki} = \sqrt{\frac{2}{\pi} J_k(\omega_{ki}) \Delta\omega}.  \qquad (7)$$

with a sampling interval $\Delta\omega$.

Similar to previous work[86], we chose 100 modes to perform such discretization. This confirms that the same Hamiltonian was employed in both quantum and quasiclassical dynamics; whether or not the current discretization is enough to reproduce the bath behavior is beyond the current purpose of this work.

Following our previous work [86], we took a rather simple approach to characterize the electron-phonon coupling strength. When the bath organization energy is represented by a single mode with a characteristic frequency $\omega_c$, the effective electron-phonon coupling strength is

$$\kappa_{eff}/\omega_c = \sqrt{\frac{2\lambda}{\omega_c}},  \qquad (8)$$

A.2. Linear vibronic coupling models

Linear vibronic coupling models have been widely used to describe the nonadiabatic dynamics at conical intersections. Here, we choose the excited-state internal conversion dynamics of pyrazine via the S1/S2 (n$\pi^*$--$\pi\pi^*$) conical intersection as the second group of typical examples[2]. The two lowest excited states of pyrazine are strongly coupled to each other, resulting in a conical intersection that opens the internal conversion channel and gives rise to very broad absorption spectra with interesting vibronic features. Pyrazine often serves as a prototype to examine the performance of newly developed dynamical methods[2, 16]. In this work, we consider two models (a 3-mode model and 24-mode model) for the S1/S2 (n$\pi^*$ -- $\pi\pi^*$) conical intersection of pyrazine[87, 88]. For both of them, the linear vibronic coupling models were taken as

$$H = T_{nuc} + \sum_{k}^{2} |\phi_k\rangle V_{kk} \langle\phi_k| + \sum_{k \neq l}^{2} |\phi_k\rangle V_{kl} \langle\phi_l|, \qquad (9)$$

where $T_{nuc}$ denotes the nuclear kinetic energy and $V_{kk}$ and $V_{kl}$ denote the diagonal and off-diagonal elements of the electronic potential matrix, which are represented as follows

$$V_{kk} = E_k + \sum_{c} \frac{1}{2}\omega_c Q_c^2 + \sum_{t} \frac{1}{2}\omega_t Q_t^2 + \sum_{t} \kappa_t^{(k)} Q_t, \qquad (10)$$

$$V_{12} = V_{21} = \sum_{c} \lambda_c Q_c, \qquad (11)$$

where $E_k$, $k = 1, 2$, are the vertical excitation energies of the n$\pi^*$ and $\pi\pi^*$ state, $Q_c$ is the coupling mode, $\lambda_c$ is the vibronic coupling constant, $Q_t$ ($t$ = 1, 2 for 3-mode model and $t$ = 1, 2, …, 23 for 24-mode model) are the tuning modes, $\omega_c$ and $\omega_t$ are the corresponding frequencies of these modes, and $\kappa_t^{(k)}$ ($k$ = 1, 2, $t$ = 1, 2 for 3-mode model and $t$ = 1, 2, …, 23 for 24-mode model) represent the gradients of the

n$\pi^*$ and $\pi\pi^*$ potentials with respect to the tuning modes. All details and parameters in these two models can be found in previous works[87, 88].

**B. Li-Miller Quasiclassical Methods**

B.1. SQC-LM dynamics

The Li-Miller mapping Hamiltonian is given as[81, 84]

$$H = \sum_{k=1}^{F} H_{nn}\left(x^{(k)}p_y^{(k)} - y^{(k)}p_x^{(k)}\right)$$
$$+ \sum_{k<k'} H_{kk'}\left[\left(x^{(k)}p_y^{(k')} - y^{(k)}p_x^{(k')}\right) + \left(x^{(k')}p_y^{(k)} - y^{(k')}p_x^{(k)}\right)\right] \quad (12)$$

In the pure classical treatment, the electronic population of the $k$-th state can be obtained by averaging the quantum occupation number $n_k$ over all trajectories[2, 84],

$$P = \langle n_k \rangle_{traj} = \langle x^{(k)}p_y^{(k)} - y^{(k)}p_x^{(k)} \rangle_{traj}, \quad (13)$$

where the bracket $\langle ... \rangle_{traj}$ refers to an average taken over all trajectories.

In the quasi-classical dynamics, it is possible to estimate the distribution of the final quantum states (occupation) by calculating the final values of the action variables $n_k$ accumulated in square histograms ("bin") centered at quantum occupation values $N = 0$ or 1. Cotton and Miller proposed that a "bin" idea should be employed for both initial-state sampling and final-state assignment, and they applied this idea in the SQC-MM method[59, 60]. We borrowed this idea and define the SQC-LM dynamics. After the inclusion of the zero-point (ZP) correction on the electronic mapping degrees of freedom [55, 56, 85], the classical LM mapping Hamiltonian becomes

$$H = \sum_{k=1}^{F} H_{kk} \left( x^{(k)} p_y^{(k)} - y^{(k)} p_x^{(k)} - \gamma \right)$$
$$+ \sum_{k<k'} H_{kk'} \left[ \left( x^{(k)} p_y^{(k')} - y^{(k)} p_x^{(k')} \right) + \left( x^{(k')} p_y^{(k)} - y^{(k')} p_x^{(k)} \right) \right] \quad (14)$$

where $\gamma$ is the ZP correction parameter.

B.2. Symmetrical Window Methods

Recently, Cotton and Miller proposed a triangle window function[66, 79]. We took this method into account in this work. For a 2-state model, such functions are defined as

$$W_1(n_1, n_2) = 2 \cdot h(n_1 + \gamma - 1) \cdot h(n_2 + \gamma) \cdot h(2 - 2\gamma - n_1 - n_2)$$
$$W_2(n_1, n_2) = 2 \cdot h(n_1 + \gamma) \cdot h(n_2 + \gamma - 1) \cdot h(2 - 2\gamma - n_1 - n_2) \quad (15)$$

where $h(z)$ is the Heaviside function, $h(z) = 0$ when $z < 0$ and $h(z) = 1$ when $z \geq 0$. Equation (15) gives two triangle windows. The parameter $\gamma$ is generally chosen to be 1/3 for the triangle window function.

The time-dependent populations $P_{map,k}$ of the electronic states are evaluated by averaging over all trajectories

$$P_{map,k}(t) = \frac{\langle W_k(\mathbf{n} = n_1, ... n_k, ... n_F) \rangle}{\sum_{k'}^{F} \langle W_{k'}(\mathbf{n} = n_1, ... n_{k'}, ... n_F) \rangle} \quad (16)$$

B.3. Initial Sampling Methods

In the initial sampling, the two coordinates and two momenta should be sampled for each angular momentum. Because only the value of angular momentum defines a constrain, different sampling methods can be employed to get different initial conditions that return to the correct the angular momentum.

B.3.1. Sampling Method I

In the SQC-LM dynamics simulations, the initial-state sampling can be performed using the action-angle sampling method,[84] namely,

$$
\begin{aligned}
x^{(k)}(0) &= \sqrt{n_k + \gamma} \cos\theta \\
y^{(k)}(0) &= -\sqrt{n_k + \gamma} \sin\theta \\
p_x^{(k)}(0) &= \sqrt{n_k + \gamma} \sin\theta \\
p_y^{(k)}(0) &= \sqrt{n_k + \gamma} \cos\theta
\end{aligned}
\quad (17)
$$

where the angle $\theta \in [-\pi, \pi)$. The distribution of $n_k$ values depend on the employed symmetrical window method. This initial sampling is suggested by Liu [84], which satisfies, $x^{(k)}(0) = p_y^{(k)}(0)$ and $y^{(k)}(0) = -p_x^{(k)}(0)$. It is important to point out that the SQC-LM dynamics with this initial sampling method is equivalent to the SQC-MM dynamics with the action-angle sampling of mapping variables. [84]

B.3.2. Sampling Method II

We notice that only constraint of the initial sampling is

$$
x^{(k)} p_y^{(k)} - y^{(k)} p_x^{(k)} = n_k + \gamma. \quad (18)
$$

Thus, it is possible to perform the initial sampling according to Eq. (22) without any additional constraints. The relevant sampling is implemented in the below procedure. We randomly choose $\tilde{x}^{(k)}$ ($\tilde{y}^{(k)}$, $\tilde{p}_x^{(k)}$ or $\tilde{p}_y^{(k)}$) $\in$ [-$\xi$, $\xi$], where $\xi$ is an arbitrary positive number. For instance, we chose $\xi =1$ here without losing generality. When $a = \tilde{x}^{(k)} \tilde{p}_y^{(k)} - \tilde{y}^{(k)} \tilde{p}_x^{(k)} > 0$, we can directly rescale the four variables to achieve the correct angular momentum value $n_k + \gamma$ according to

$$
\begin{aligned}
x^{(k)}(0) &= \sqrt{n_k + \gamma/a}\, \tilde{x}^{(k)} \\
y^{(k)}(0) &= \sqrt{n_k + \gamma/a}\, \tilde{y}^{(k)} \\
p_x^{(k)}(0) &= \sqrt{n_k + \gamma/a}\, \tilde{p}_x^{(k)} \\
p_y^{(k)}(0) &= \sqrt{n_k + \gamma/a}\, \tilde{p}_y^{(k)}
\end{aligned} \qquad (19)
$$

When $a = \tilde{x}^{(k)} \tilde{p}_y^{(k)} - \tilde{y}^{(k)} \tilde{p}_x^{(k)} < 0$, we simply exchange the values $\tilde{x}^{(k)}$ and $\tilde{y}^{(k)}$ as well as the values of $\tilde{p}_x^{(k)}$ and $\tilde{p}_y^{(k)}$ to make $a = \tilde{x}^{(k)} \tilde{p}_y^{(k)} - \tilde{y}^{(k)} \tilde{p}_x^{(k)} > 0$. Then the above algorithm in Eq. (19) can be employed again to get the initial sampling.

B.3.3. Sampling Method III

By adding constraints, $x^{(k)} p_y^{(k)} > 0$ and $y^{(k)} p_x^{(k)} < 0$ after the consideration of Eq. (18), we obtain Sampling Method III. When sampling $\tilde{x}^{(k)}$, $\tilde{y}^{(k)}$, $\tilde{p}_x^{(k)}$ and $\tilde{p}_y^{(k)}$, we keep the same sign for $x^{(k)}$ and $p_y^{(k)}$, and the different signs for $\tilde{y}^{(k)}$ and $\tilde{p}_x^{(k)}$. This gives $a = \tilde{x}^{(k)} \tilde{p}_y^{(k)} - \tilde{y}^{(k)} \tilde{p}_x^{(k)} > 0$. Next, we may employ Eq. (19) to get the initial conditions.

B.4. Computational details

For site-exciton models [68, 86], the initial condition is generated by putting the lowest vibrational level of the ground electronic state into the donor state, according to the Condon approximation.

For the 3-mode and 24-mode linear vibronic coupling models for pyrazine, the initial condition is generated by putting the lowest vibrational level of the ground electronic state into the ππ* state.

For the nuclear part, Wigner sampling was used for initial sampling. A symplectic algorithm[84] was employed to propagate the electronic part and nuclear part of a trajectory. For the nuclear part, the symplectic algorithm is equivalent to the Velocity Verlet algorithm[93]. The electronic integration time step was 0.001 fs and the nuclear integration time step 0.01 fs. For each model, 5000 trajectories were used in the calculation.

**C. ML-MCTDH**

The MCTDH method [4] and its extension, the ML-MCTDH method [5, 94, 95], are numerically accurate full quantum dynamics methods. Discussions on their theoretical frameworks and simulation tricks can be found in previous works. [5, 89, 92, 94-97] Similar to our previous work [86], we employed these methods to benchmark the accuracy of the LM dynamics. The Heidelberg MCTDH package [98] was used to perform the ML-MCTDH calculations.

## III. RESULTS AND DISCUSSION

### A. SQC-LM dynamics in symmetric site-exciton models

The first test set of models represent the symmetric site-exciton models with $\Delta E = 0$ and $V_{12} = 0.0124$ eV, which displays a Rabi oscillation frequency of ~200 cm$^{-1}$ in the pure electronic dynamics. The dynamic results are collected in Fig. 1.

In the case of $\omega_c = 200$ cm$^{-1}$, the SQC-LM method with all initial conditions gives the reasonable results compared with ML-MCTDH method [86], no matter whether the system-bath coupling is weak or strong. However, the weaker population oscillation amplitudes are obtained by the SQC-LM dynamics with Sampling Method II and III. As the contrast, the SQC-LM dynamics using Sampling Method I (equivalent to the SQC-MM dynamics with the action-angle sampling of mapping variables) shows the good agreement with the ML-MCTDH dynamics. Overall, among three sampling method, Sampling Method I is the best selection in the SQC-LM dynamics in the description of the current site-exciton model.

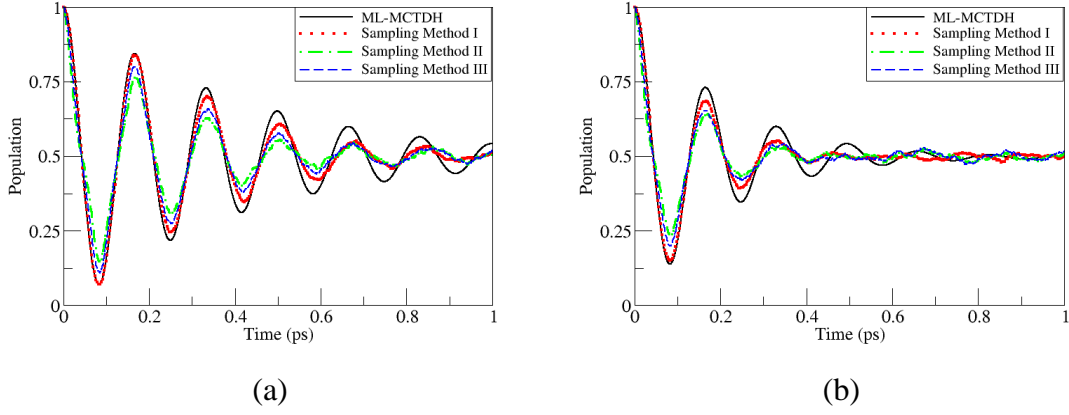

(a)                                      (b)

Fig. 1. Time-dependent electronic population of the donor state in various two-state models with $\Delta E = 0$ and $V_{12} = 0.0124$ eV. (a) $\omega_c = 200$ cm$^{-1}$, $\kappa_{eff}/\omega_c = 0.5$; (b) $\omega_c = 200$ cm$^{-1}$, $\kappa_{eff}/\omega_c = 0.7$. The ML-MCTDH results are taken from our previous work.[86]

**B. SQC-LM dynamics in asymmetric site-exciton models**

For the asymmetric site-exciton models with $\Delta E = V_{12} = 0.0124$ eV (Rabi frequency ~ 224 cm$^{-1}$) and $\omega_c = 200$ cm$^{-1}$, the SQC-LM dynamics with the Sampling Method I (equivalent to the SQC-MM dynamics) almost reproduces the exact results provided by the ML-MCTDH approach, see Fig. 2. The results based on SQC-LM dynamics using Sampling Method II and III are still satisfactory, while some minor deviation exists and the population oscillation become slighter weaker.

For both of symmetric and asymmetric site-exciton models, the SQC-LM dynamics basically provides the acceptable results while the results are also dependent on the initial sampling. Sampling Method II and III seem not improve the performance of the SQC-LM dynamics. On the contrary, they gave slightly worse results than Sampling Method I. This indicates that the SQC-LM dynamics with the Sampling Method I (equivalent to the SQC-MM dynamics with the action-angle sampling of mapping variables) should be chosen for the study of similar systems.

In the current work, we do not provide the results of site-exciton model with high-frequency baths. It is clearly known that the treatment of fast bath modes with pure classical dynamics is not correct. However, such problems may be solved by using adiabatic renormalization or relevant approaches [86, 99, 100].

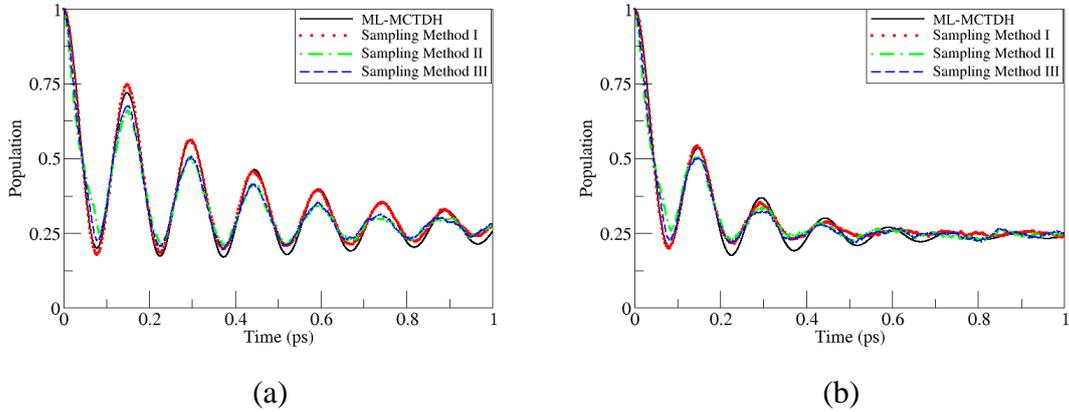

(a)                                  (b)

Fig. 2. Time-dependent electronic population of the donor state in various two-state models with $\Delta E = 0.0124$ eV and $V_{12} = 0.0124$ eV. (a) $\omega_c = 200$ cm$^{-1}$, $\kappa_{eff}/\omega_c = 0.5$; (b) $\omega_c = 200$ cm$^{-1}$, $\kappa_{eff}/\omega_c = 0.7$. The ML-MCTDH results are taken from our previous work.[86]

## C. Linear vibrionic coupling models for conical intersection

Next, the linear vibronic coupling models for the nonadiabatic dynamics of pyrazine are also considered as typical models. In the ML-MCTDH dynamics of the 3-mode Hamiltonian model with one coupling and two tuning modes, the ππ* population decay takes place very quickly, while strong recurrence is observed. Later, the population tends to become stable, see Fig. 3.

The SQC-LM dynamics with Sampling Method I (equivalent to the SQC-MM dynamics) provides fairly reasonable results compared to those of ML-MCTDH for 3-mode model of pyrazine, while the ππ* state population becomes smaller than that of ML-MCTDH after about 150 fs [Fig. 3(a) and Fig. 3(c)]. This implies that the SQC-LM with Sampling Method I and ML-MCTDH methods do not give the same asymptotic limit and their difference is about 0.1. When other sampling approaches, Sampling Method II and III, were employed, the SQC-LM dynamics gives the asymptotical limits closer to that of ML-MCTDH.

When 24-mode model for pyrazine is considered, the overall population decay becomes faster in the ML-MCTDH dynamics and the recurrence becomes much weaker [Fig. 3(b) and Fig. 3(d)]. This is because the presence of many modes causes dissipative effects in the nonadiabatic dynamics. The SQC-LM methods with the Sampling Method II and III still show slightly better results than the SQC-LM dynamics with Sampling Method I, as shown in Fig. 3(b) and 3(d). Particularly, when Sampling Method II is employed, the SQC-LM dynamics seems to be correctly capture the long-time dynamics features compared with the ML-MCTDH results. We also noticed that this sampling approach also improves the description of the oscillation patterns in the time-dependent population. This indicates that the performance of the SQC-LM dynamics may be improved by the proper selection of different initial conditions, when the conical intersection models are considered. In the current conical intersection model, Sampling Method II is the best choice.

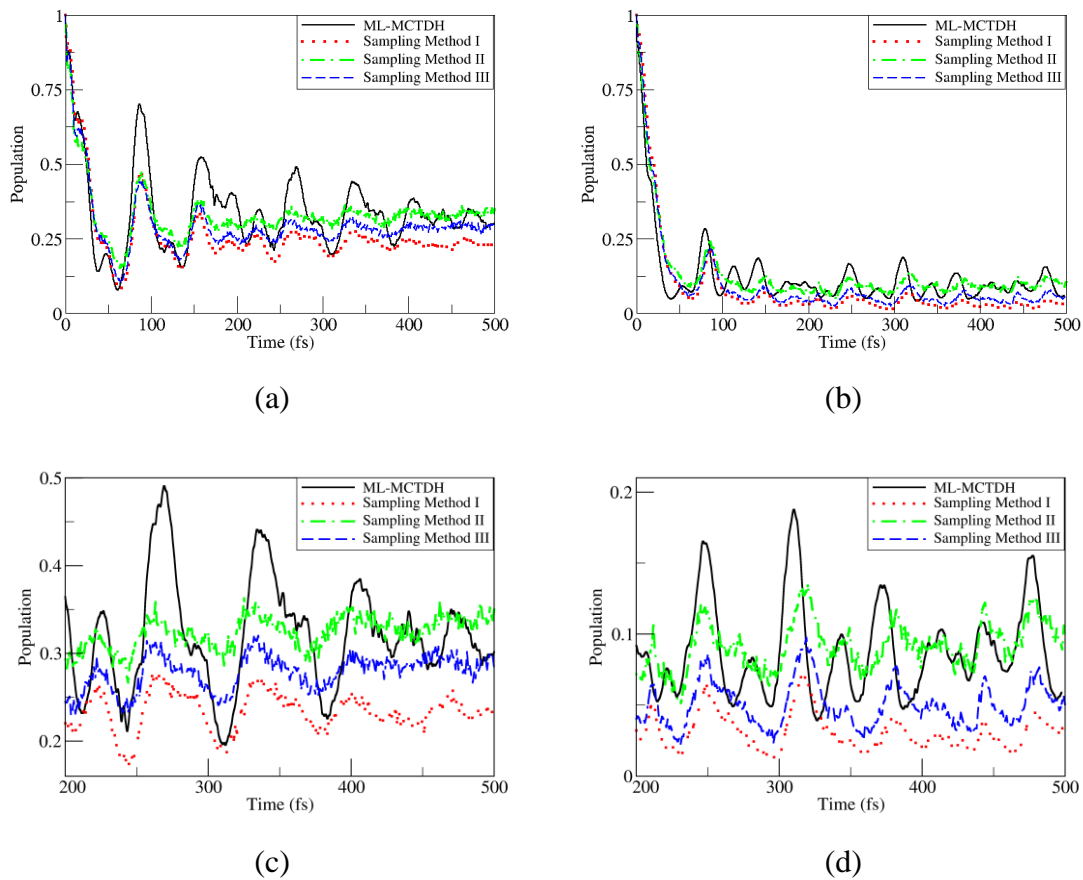

Fig. 3. Time-dependent electronic population of the ππ* state in pyrazine models. For (a) and (c), the model is 3-mode model; For (b) and (d), the model is 24-mode model. The population from 200 to 500 fs is given in (c) and (d) with different population-axis range. The ML-MCTDH results are calculated by ourselves, which are consistent with previous results.[87, 88]

It is interesting to notice that two test models in fact show reversed conclusions in the discussion of the influence of the different initial sampling on the SQC-LM dynamics. In the site-exciton model, Sampling Method I that is defined by using the additional constraint [Eq. (17)] condition gives the best performance of the SQC-LM dynamics. This also means that the SQC-MM dynamics with action-angle sampling of the mapping variables should behave equally well. However, Sampling Method II that is the most flexible sampling approach without any additional constraints seems to give the best results in the conical intersection models. This highly indicates that the suitable initial conditions may improve the performance of the SQC-LM dynamics with the proper selection of the initial condition is system dependent.

## IV. CONCLUSIONS

In this paper, we evaluated the performance of the SQC-LM dynamics in the treatment of nonadiabatic dynamics. We considered the window technology and the ZP correction for the electronic mapping variables in the SQC-LM dynamics. The SQC-LM approach allows the flexible sampling techniques since 4 variables, instead of 2 variables in the SQC-MM model, are available for 1 quantum state. Thus, we try three sampling ways, namely Sampling Method I, II and III, to generate the initial conditions and check the performance of the SQC-LM method with them. Sampling Method I is defined by the strong additional condition with $x^{(k)}(0) = p_y^{(k)}(0)$ and $y^{(k)}(0) = -p_x^{(k)}(0)$. Under this particular initial sampling, the SQC-LM dynamics and the SQC-MM dynamics with action-angle sampling of the mapping variables is equivalent. Sampling Method II, do not involve any additional constraints except the value of angular momentum. Sampling Method III lies between them and the constraints on the sign for the sampling variables are considered.

Two types of models are considered to evaluate the efficiency and accuracy of the SQC-LM dynamics method. The first group includes site-exciton models and the second group includes the typical linear vibronic coupling models for the description of conical intersections of pyrazine. In these models, our proposed initial sampling approaches show different performances. When the bath frequency is not very high, in principle the SQC-LM dynamics with Sampling Method I (equivalent to the SQC-MM dynamics) almost reproduces the exact results by ML-MCTDH. With the new sampling tricks, the SQC-LM dynamics still gives the reasonable results while the oscillation in the population dynamics becomes slightly weaker. As the contrast, the other sampling approaches gives the better results in the nonadiabatic dynamics at the conical intersections of pyrazine. In this case, the asymptotical limit in the population dynamics can be captured by the SQC-LM method with Sampling Method II and the description of the oscillation patterns also becomes better. In this sense, the SQC-LM method outperforms the SQC-LM dynamics with Sampling Method I (equivalent to the SQC-MM dynamics) in the linear vibronic coupling model for the conical intersection of pyrazine, if the proper initial condition is employed,

Overall, it is clear that the results of the SQC-LM dynamics may be relevant to the different choice of the initial conditions. Because the relation between the population and two pairs of coordinates/momenta defines the single constraint to perform the initial sampling, this implies that it is possible to develop other more flexible sampling ways to achieve the better performance of the SQC-LM dynamics. In this sense, the SQC-LM approach and LM mapping Hamiltonian are very promising in the description of nonadiabatic dynamics. At the same time, it is certainly interesting to combine other semiclassical dynamics approaches and the LM mapping Hamiltonian for the development of the novel nonadiabatic dynamics approaches.

## CONFLICTS OF INTEREST

There are no conflicts to declare.

## ACKNOWLEDGMENTS

This work is supported by NSFC projects (Nos. 21673266 and 21873112). J. Zheng thanks the support from the Natural Science Foundation of Shandong Province (ZR2018BB043) and the Postdoctoral Scientific Research Foundation of Qingdao (2017012). The authors thank the Supercomputing Center, Computer Network Information Center, CAS; National Supercomputing Center in Shenzhen and National Supercomputing Center in Guangzhou.

## APPENDIX I. Li-Miller Model

The Hamiltonian operator $\hat{H}$ of the $N$-state system is expressed as

$$\hat{H} = \sum_{k,k'} \hat{h}_{kk'} |\phi_k\rangle\langle\phi_{k'}|, \tag{20}$$

where ($\phi_k$ or $\phi_{k'}$) is the $k$-th (or $k'$-th) quantum state of the system. The diagonal element of $\hat{h}_{kk'}(k = k')$ includes the nuclear kinetic energy and potential energy on the $k$-th electronic state, while the off-diagonal element $\hat{h}_{kk'}(k \neq k')$ denotes the interstate couplings between the $k$-th and $k'$-th states.

The bra-ket form is given by a pair of creation and annihilation operators

$$\begin{aligned} |k\rangle\langle k| &= \hat{a}_k^+ \hat{a}_k \\ |k\rangle\langle k'| &= \hat{a}_k^+ \hat{a}_{k'} \end{aligned}. \tag{21}$$

Three angular momentum operators are defined as

$$\begin{aligned} \hat{\sigma}_x^{(k)} &= \hat{a}_k + \hat{a}_k^+ \\ \hat{\sigma}_y^{(k)} &= \frac{\hat{a}_k - \hat{a}_k^+}{i} \\ \hat{\sigma}_z^{(k)} &= \left[\hat{a}_k^+, \hat{a}_k\right] \end{aligned}. \tag{22}$$

The commutation relation between them is

$$\frac{i}{2}\left[\hat{\sigma}_a^{(k)}, \hat{\sigma}_b^{(k)}\right] = \varepsilon_{abc}\hat{\sigma}_c^{(k)}, \tag{23}$$

where the Levi-Civita symbol $\varepsilon_{abc}$ is equal to 1 for cyclic permutations of *xyz*, -1 for anti-cyclic permutations, and 0 for $a = b$. Considering Eqs. (21), (22) and (23), the *N*-state Hamiltonian in Eq. (20) becomes

$$\hat{H} = \sum_k \frac{1}{2}\left(\hat{\mathbf{1}}^{(k)} + \frac{i}{2}\left[\hat{\sigma}_x^{(k)}, \hat{\sigma}_y^{(k)}\right]\right) H_{kk} + \sum_{k<k'} \frac{1}{2}\left(i\left[\hat{\sigma}_x^{(k)}, \hat{\sigma}_y^{(k')}\right] - i\left[\hat{\sigma}_y^{(k)}, \hat{\sigma}_x^{(k')}\right]\right) H_{kk'}. \tag{24}$$

Then, the quantum angular momentum is remapped to its classical counterpart $\mathbf{L} = \mathbf{x} \times \mathbf{p}$, and we have

$$x_a p_b - x_b p_a = \varepsilon_{abc} L_c, \tag{25}$$

where $\{x_a, x_b, x_b\} = \{x, y, z\}$ and $\{p_a, p_b, p_b\} = \{p_x, p_y, p_z\}$. Finally, this gives the Li-Miller mapping Hamiltonian [Eq. (12)]. [81, 84] Taking ZP correction into account, Eq. (14) is obtained. The equations of motion can be obtained as

$$\begin{aligned}
\dot{x}^{(k)} &= \frac{\partial H}{\partial p_x^{(k)}} = -H_{kk} y^{(k)} - \sum_{k' \neq k} H_{kk'} y^{(k')} \\
\dot{y}^{(k)} &= \frac{\partial H}{\partial p_y^{(k)}} = H_{kk} x^{(k)} + \sum_{k' \neq k} H_{kk'} x^{(k')} \\
\dot{p}_x^{(k)} &= -\frac{\partial H}{\partial x^{(k)}} = -H_{kk} p_y^{(k)} - \sum_{k' \neq k} H_{kk'} p_y^{(k')} \\
\dot{p}_y^{(k)} &= -\frac{\partial H}{\partial y^{(k)}} = H_{kk} p_x^{(k)} + \sum_{k' \neq k} H_{kk'} p_x^{(k')}
\end{aligned} \quad (26)$$

## APPENDIX II. Further Discussion on Initial Sampling

Liu[84] proposed to use action-angle method to sample the initial coordinates and momenta of the electronic part, which is Sampling Method I in the current work in which the ZP correction and window trick are considered. Actually, combining this initial sampling, $x^{(k)}(0) = p_y^{(k)}(0)$ and $y^{(k)}(0) = -p_x^{(k)}(0)$, with the equations of motion [Eq. (26)], we can obtain $x^{(k)}(t) = p_y^{(k)}(t)$ and $y^{(k)}(t) = -p_x^{(k)}(t)$ at any time $t$. This goes back to quasiclassical dynamics based on the MM model. However, when the symplectic method is employed to propagate the classical trajectory in the implementation of the LM dynamics, the coordinates and momentum components are updated in the two successive steps. Thus; after some propagation time, the $x^{(k)}(t) = p_y^{(k)}(t)$ and $y^{(k)}(t) = -p_x^{(k)}(t)$ may not always be satisfied with time being.

In this case, some deviation may appear in the long-time single-trajectory propagation for the LM dynamics and MM dynamics.

It is also important to figure out that this initial sampling always gives

$$\left(x^{(k)}(t)\right)^2 + \left(y^{(k)}(t)\right)^2 = \left(p_x^{(k)}(t)\right)^2 + \left(p_y^{(k)}(t)\right)^2. \tag{27}$$

Thus, the length of the coordinate vector is always the same as the length of the momentum vector, i.e., $|\mathbf{x}| = |\mathbf{p}|$. If the interstate coupling is not considered, the trajectory will propagate along a circle in the $x$-$y$ plane with the radius $|\mathbf{x}| = \sqrt{\left(x^{(k)}(t)\right)^2 + \left(y^{(k)}(t)\right)^2}$.

If we assume $|\mathbf{x}| = \beta |\mathbf{p}|$, it is also possible to show that the classical LM dynamics with this initial sampling is also equivalent to the classical MM dynamics. When $\beta = 1$, we go back to the above special situation with $|\mathbf{x}| = |\mathbf{p}|$.

When different parameters for different quantum states are used, i.e.,

$$\begin{aligned} x^{(k)}(0) &= \alpha^{(k)}(0) p_y^{(k)}(0) \\ y^{(k)}(0) &= -\beta^{(k)}(0) p_x^{(k)}(0) \end{aligned}, \tag{28}$$

where $\alpha^{(k)}(0) \neq \beta^{(k)}(0)$. Then the classical LM model is different to the classical MM model.

If there is no any additional restriction on the sign of $\alpha^{(k)}(0)$ and $\beta^{(k)}(0)$, we get Sampling Method II. When we restricted $\alpha^{(k)}(0) > 0$ and $\beta^{(k)}(0) > 0$, Sampling Method III is achieved.